\documentclass[11pt]{article}
\usepackage[utf8]{inputenc}
\usepackage{listings}
\usepackage{color}
\usepackage{longtable}
\usepackage{float}
\usepackage{amsmath}
\usepackage{amssymb}
\usepackage{authblk}

\newcommand{\pf}{p_{\text{free}}}
\newcommand{\pa}{p_{\text{abs}}}
\newcommand{\pr}{p_{\text{ref}}}
\newcommand{\pck}{p_{\text{rad}}}
\newcommand{\pd}{p_{\delta}}

\newcommand{\argD}{(x,t\vert x_{0})}
\newcommand{\pfL}{\tilde{p}_{\text{free}}}
\newcommand{\paL}{\tilde{p}_{\text{abs}}}
\newcommand{\prL}{\tilde{p}_{\text{ref}}}
\newcommand{\pckL}{\tilde{p}_{\text{rad}}}
\newcommand{\pdL}{\tilde{p}_{\delta}}
\newcommand{\argL}{(x,s\vert x_{0})}

\newcommand{\dplL}{\frac{\partial\paL (\xi,s\vert x_{0})}{\partial\xi}\bigg\vert_{\xi=a}}
\newcommand{\dprL}{\frac{\partial\paL (x,s\vert \xi)}{\partial\xi}\bigg\vert_{\xi=a}}
\newcommand{\dpflL}{\frac{\partial\pfL (\xi,s\vert x_{0})}{\partial\xi}\bigg\vert_{\xi=a}}

\title{}
\date{\today}
\begin{document}

\title{Path integral approach to theories of diffusion-influenced reactions}
\author{Thorsten Pr\"ustel} 
\author{Martin Meier-Schellersheim} 
\affil{Laboratory of Systems Biology\\National Institute of Allergy and Infectious Diseases\\National Institutes of Health}
\maketitle
\let\oldthefootnote\thefootnote 
\renewcommand{\thefootnote}{\fnsymbol{footnote}} 
\footnotetext[1]{Email: prustelt@niaid.nih.gov, mms@niaid.nih.gov} 
\let\thefootnote\oldthefootnote
\begin{abstract}
The path decomposition expansion represents the propagator of the irreversible reaction as a convolution of the first-passage, last-passage and rebinding time probability densities. Using path integral technique, we give an elementary, yet rigorous, proof of the path decomposition expansion of the Green's functions describing the non-reactive case and the irreversible reaction of an isolated pair of molecules. To this end, we exploit the connection between boundary value problems and interaction potential problems with $\delta$- and $\delta'$-function perturbation. In particular, we employ a known exact summation of a perturbation series to derive exact relations between the Green's functions of the perturbed and unperturbed problem. Along the way, we are able to derive a number of additional exact identities that relate the propagators describing the free-space, the non-reactive as well as the completely and partially reactive case.
\end{abstract}
\section{Introduction}
\label{sec-1} 
The transition probability density function (PDF) $p(x,t\vert x_{0})$ is the fundamental dynamical quantity that describes the diffusion-influenced reaction of an isolated pair of molecules. Traditionally, $p(x,t\vert x_{0})$ appears as Green's function (GF) of the diffusion equation supplemented by suitable boundary conditions (BC) imposed at the encounter distance $x = a$ that implement the actual reaction \cite{smoluchowski:1917, collins1949diffusion, Goodrich:1954, Goesele:1984, Rice:1985, Agmon:1984, kimShin:1999, TPMMS_2012JCP}.
However, imposing BC are not the only way to implement reactions. Alternatively, sink terms may be added to the diffusion equation. It has been discussed that, under certain conditions, sink terms are equivalent to a description in terms of BC, but they provide a more general and more flexible approach than the traditional one \cite{Wilemski:1973, Doi_1:1976, Doi_2:1976, North-1978, Khokhlova:2012BullKorCS, Prustel_Area:2014, Prustel_Area_General:2014, Isaacson:2014}. Also, they offer another adavantage: The diffusion equation with sink terms assumes a form that is quite similar to the Schroedinger equation. In fact, it is well-known that both equations are related by a Wick-transformation, that is by $t \rightarrow -it$ \cite{Roepstorff:1994, Chaichian:2001, Schulman:2005}. In this way, methods that have been used in a quantum-mechanical context, can be borrowed to study the theory of diffusion-influenced reactions. A case in point is the path decomposition expansion (PDX), which was originally employed to study quantum-mechanical problems \cite{Auerbach:1985, Halliwell:1993, Baal-1991, Schulman:2005, Yearsley:2008, Yearsley:2009}. Recently, it has been shown that in the context of diffusion-influenced reactions the GF of the irreversible reaction can be factorized in the Laplace domain as a product of the first-, last-passage and rebinding time probability densities \cite{Prustel-qm-2015}. In this manuscript, we present a straightforward, yet rigorous, proof of the PDX of the propagators that play an important role in the theory of diffusion-influenced reactions. The proof is based on another technique imported from quantum mechanics. In a series of papers \cite{Grosche:1990, Grosche-PRL-1993, Grosche-1995}, Grosche discussed the implementation of Dirichlet and Neumann BC in terms of interaction potentials that are $\delta$- and $\delta'$-function perturbed. By using an exact summation of the expansion of the corresponding path integral, an exact convolution relation between the perturbed and unperturbed GF can be established. Because Dirichlet and Neumann BC correspond to absorbing (completely reactive) and reflecting (non-reactive) BC in the theory of diffusion-influenced reactions, the Wick-rotated (Euclidean) version of Grosche's results can be immediately applied to the case of an isolated pair. By considering different choices for the unperturbed potential, we can derive several important relations between the GF corresponding to various BC, which are far from obvious from a differential equation point of view \cite{Agmon-1991}. With these relations at our disposal, the actual proof of the PDX becomes rather accessible.
We point out that some of the relations that we will derive have been obtained before by using the general form of the GF of the Smoluchowski equation \cite{Agmon-1991}. Alternative proofs of the PDX have been given in Refs.~\cite{Auerbach:1985, Halliwell:1993, Baal-1991, Halliwell-1995}.
The application of quantum field theoretical methods to diffusion-limited reactions and path integral techniques to stochastic differential equation are reviewed in Ref.~\cite{Mattis-1998} and \cite{Chow-2015}, respectively. 
The manuscript is structured as follows. First, we will briefly describe the traditional BC approach and introduce the alternative formulation given in terms of sink potentials. Then, we will briefly recall the PDX and summarize Grosche's results. Equipped with those, we will derive various convolution relations, from which the PDX results.
In the following, we will consider an isolated pair in one dimension (1D) for convenience. We will study the 2D and 3D cases in a forthcoming manuscript.
\section{Theory}
\subsection{Boundary vs potential problems}
The conditional PDF $p(x,t\vert x_{0})$, which yields the probability to find the molecule at position $x$ at time $t$, given that it was initially localized at $x_{0}$, is the GF of the diffusion equation 
\begin{equation}
\frac{\partial p(x,t\vert x_{0})}{\partial t} = D \frac{\partial^{2} p(x,t\vert x_{0})}{\partial x^{2}},
\end{equation}
subject to the initial condition (IC) 
\begin{equation}
p(x,t=0\vert x_{0}) = \delta(x-x_{0}).
\end{equation}
The traditional, Smoluchowski-Collins-Kimball (SCK) approach \cite{smoluchowski:1917, collins1949diffusion, Goesele:1984, Rice:1985} to incorporate the actual binding reaction, amounts to impose a BC at the encounter distance $x=a$. The Collins-Kimball, or radiation BC reads \cite{collins1949diffusion, Goesele:1984, Rice:1985, Agmon:1984, kimShin:1999, TPMMS_2012JCP}
\begin{equation}
D\frac{\partial p(x, t\vert x_{0})}{\partial x}\vert_{x=a} = \kappa_{a}p(x=a,t\vert x_{0}).
\end{equation}
The absorbing BC is recovered for $\kappa_{a} \rightarrow \infty$, while the non-reactive, reflective BC is obtained for $\kappa_{a} = 0$. We will denote the GF corresponding to free-space, absorbing, reflecting and radiation BC as $\pf\argD$, $\pa\argD$, $\pr\argD$ and $\pck\argD$, respectively.

Instead of imposing BC at the encounter distance, the actual binding reaction may alternatively be taken into account by introducing sink terms in the diffusion equation. For instance, under certain conditions, the radiation BC may be recovered from the diffusion equation with an additional $\delta$-function interaction potential \cite{Wilemski:1973}
\begin{equation}\label{Bloch-Equation}
\frac{\partial p(x,t\vert x_{0})}{\partial x} = D \frac{\partial^{2} p(x,t\vert x_{0})}{\partial x^{2}} - \gamma \delta(x-a).
\end{equation}
Note that Eq.~(\ref{Bloch-Equation}) does not necessarily come with any BC imposed at the encounter distance. Also, we emphasize that the solution of Eq.~(\ref{Bloch-Equation}) is not identical to the solution of the radiation BC problem \cite{Taitelbaum:1992}, unless one makes additional assumptions \cite{Wilemski:1973}. One obvious difference is that the GF solution to Eq.~(\ref{Bloch-Equation}) is also defined for $x < a$ and $x_{0} < a$, in contrast to $\pck\argD$. We will use superscripts $\pd^{++}$ to refer to the solution of Eq.~(\ref{Bloch-Equation}) for $x, x_{0} > a$. Note that even $\pd^{++}$ is not equal to $\pck$, although both are defined on the same domain $x, x_{0} > a$ \cite{Taitelbaum:1992}.
\subsection{Path decomposition expansion}
The PDX \cite{Auerbach:1985, Halliwell:1993, Schulman:2005} allows to break space into different regions and to study the individual contributions from the propagators restricted to one region to the complete propagator across all regions. While the decomposition of space may be chosen arbitrarily, the existence of a non-vanishing potential may suggest a natural decomposition. This is in particular true for a $\delta$- and Heaviside step-function potential. These potentials are of special interest in the theory of diffusion-influenced reactions, because they correspond to the SCK and Doi model, respectively \cite{Doi_1:1976, Doi_2:1976, Khokhlova:2012BullKorCS, Prustel_Area_General:2014, Isaacson:2014}.
It has been shown by explicit calculation that the PDX indeed is satisfied by the associated propagators (in 1D and 2D) \cite{Prustel-qm-2015}. Also, it was pointed out that in the context of theories of diffusion-influenced reactions, the PDX corresponds to the convolution of three prominent time PDF: the first- and last-passage as well as the rebinding time PDF \cite{Prustel-qm-2015}. Furthermore, the PDX has been employed in a stochastic simulation algorithm that does not require explicit analytical representations of the exact GF to propagate an isolated pair correctly \cite{Prustel-sim-2015}.
Here, we shall prove the following form of the PDX
\begin{equation}
\tilde{p}(x,s\vert x_{0}) = \paL\argL + D^{2}\dprL \tilde{p}(a,s\vert a) \dplL, 
\end{equation}
where $\tilde{p}(x,s\vert x_{0})$ refers either to $\pdL^{++}(x,s\vert x_{0})$, $\prL\argL$ or $\pckL\argL$.
The proof we shall present here is based on an exactly summable perturbative expansion of a path integral that gives the propagator associated with $\delta$- and $\delta^{\prime}$-function perturbed interaction potentials. In the following, we will always consider $x, x_{0} \geq a$. Also, without loss of generality, we may assume that $a=0$.
\subsection{Path integral representation of $\delta$-function perturbed potentials}
Following Grosche \cite{Grosche:1990, Grosche-PRL-1993, Grosche-1995}, we consider a potential $W(x)$ comprised of an arbitrary potential $V(x)$ that is perturbed by a $\delta$-function term:
\begin{equation}\label{Def-W}
W(x) \equiv V(x) + \mathcal{V}(x) \equiv V(x) - \gamma \delta(x-a).
\end{equation}
The corresponding GF enjoys the following path integral representation
\begin{equation}
p_{W}(x, t\vert x_{0}) = \int \mathcal{D}x(\tau) e^{-\int^{t}_{0}[\frac{1}{4D}\dot{x}^{2}(\tau) + W(x)]d\tau},
\end{equation}
while the GF for the unperturbed problem is given by
\begin{equation}
p_{V}(x, t\vert x_{0}) = \int \mathcal{D}x(\tau) e^{-\int^{t}_{0}[\frac{1}{4D}\dot{x}^{2}(\tau) + V(x)]d\tau}.
\end{equation}
The path integral can be expanded in a perturbation series as follows
\begin{eqnarray}
p_{W}(x, t\vert x_{0}) &=& p_{V}(x, t\vert x_{0}) + \sum^{\infty}_{n=1}\gamma^{n} \int^{t}_{0}d\tau^{(n)}\int^{\tau^{(n)}}_{0}d\tau^{(n-1)}\ldots \int^{\tau^{(2)}}_{0}d\tau^{(1)}\nonumber\\
&\times& p_{V}(a, \tau^{(1)}\vert x_{0}) \ldots  p_{V}(a, \tau^{(n)}-\tau^{(n-1)}\vert a) p_{V}(x, t-\tau^{(n)}\vert a). 
\end{eqnarray}
The form of this expansion suggests to use the Laplace transformed GF $\tilde{p}(x, s\vert x_{0})$, which is defined by
\begin{equation}
\tilde{p}(x, s\vert x_{0}) = \int^{\infty}_{0} e^{-st}p(x, t\vert x_{0})dt.
\end{equation}
Then, by virtue of the convolution theorem of the Laplace tranform, one obtains in the Laplace domain
\begin{eqnarray}\label{W-V-GF}
\tilde{p}_{W}(x, s\vert x_{0}) &=& \tilde{p}_{V}(x, s\vert x_{0}) + \sum^{\infty}_{n=1}\gamma^{n} \tilde{p}_{V}(x, s\vert a)\tilde{p}_{V}(a, s\vert x_{0})[\tilde{p}_{V}(a, s\vert a)]^{n-1} \nonumber\\
&=&\tilde{p}_{V}(x, s\vert x_{0}) + \gamma\frac{\tilde{p}_{V}(x, s\vert a)\tilde{p}_{V}(a, s\vert x_{0})}{1-\gamma\tilde{p}_{V}(a, s\vert a)}.
\end{eqnarray}
As we will see, this convolution relation is sufficient to prove the PDX relation by suitable choice of $V(x)$ and $W(x)$ and by taking certain limits. For instance, the limit of an infinitely repulsive $\delta$-perturbation $\gamma\rightarrow-\infty$ implements a Dirichlet, that is, absorbing BC \cite{Grosche:1990, Grosche-PRL-1993}.  
\subsection{Proof of the PDX}
As a prerequisite, we shall derive convolution relations that connect $\pfL$ with $\paL$ and $\pdL^{++}$. 
We point out that $\pf\argD$, $\pa\argD$, and $\pd^{++}\argD$ correspond to the potentials $V(x) = 0, \gamma=0$, $V(x) = 0, \gamma\rightarrow-\infty$ and $V(x) = 0, -\infty<\gamma < 0$, respectively (see Eq.~(\ref{Def-W})).
\subsubsection*{Relation between $\pdL^{++}$ and $\pfL$}
We consider first the free-particle case, that is $V(x) = 0$, perturbed by $\mathcal{V}(x)= -\gamma\delta(x-a),  -\infty < \gamma < 0$. Then, by virtue of Eq.~(\ref{W-V-GF}) it immediately follows that
\begin{eqnarray}\label{conv_pd_pa}
\pdL^{++}(x, s\vert x_{0})=\pfL(x, s\vert x_{0}) + \gamma\frac{\pfL(x, s\vert a)\pfL(a, s\vert x_{0})}{1-\gamma\pfL(a, s\vert a)}.
\end{eqnarray}
As a consequence of Eq.~(\ref{conv_pd_pa}), we arrive at
\begin{equation}\label{ratio_pd_pf}
\frac{\pdL^{++}(a,s\vert x_{0})}{\pdL^{++}(a,s\vert a)} = \frac{\pfL(a,s\vert x_{0})}{\pfL(a,s\vert a)}.
\end{equation}
\subsubsection*{Relation between $\paL$ and $\pfL$}
If we, in addition, consider the limiting case of an infinitely repulsive $\delta$-function perturbation, that is $\gamma\rightarrow-\infty$, we obtain
\begin{equation}\label{conv_pa_pf}
\paL\argL = \pfL\argL - \frac{\pfL(x,s\vert a)\pfL(a,s\vert x_{0})}{\pfL(a, s\vert a)}.
\end{equation}
Clearly, Eq.~(\ref{conv_pa_pf}) leads to
\begin{equation}
\dplL = \dpflL - \frac{\partial\pfL(\xi,s\vert a)}{\partial\xi}\bigg\vert_{\xi\rightarrow a}\frac{\pfL(a,s\vert x_{0})}{\pfL(a,s\vert a)}.
\end{equation}
Combining this equation with an identity that results from properties of the free-space GF only (Eq.~(\ref{fundamental-free})), we obtain the relations
\begin{eqnarray}\label{derivative-pabs-pfree}
D\dplL &=& \frac{\pfL(a,s\vert x_{0})}{\pfL(a,s\vert a)} = \frac{D \partial_{\xi}\pfL(\xi,s\vert x_{0})\vert_{\xi=a}}{1+D\partial_{\xi}\pfL(\xi,s\vert a)\vert_{\xi\rightarrow a}} \nonumber \\
& = & \frac{\partial_{\xi}\pfL(\xi,s\vert x_{0})\vert_{\xi=a}}{\partial_{\xi}\pfL(\xi,s\vert x_{0}\rightarrow a)\vert_{\xi=a}},
\end{eqnarray}
where we introduced the notation $\partial_{\xi}\equiv\partial/\partial\xi$. 
Eq.~(\ref{derivative-pabs-pfree}) gives rise to
\begin{equation}\label{second-derivative-pabs-pfree}
D\frac{\partial^{2}\paL(\xi,s\vert \phi)}{\partial\xi\partial\phi}\bigg\vert_{\xi = \phi = a} = \frac{\partial_{\xi}\partial_{\phi}\pfL(\xi,s\vert \phi)\vert_{\xi=\phi=a}}{\partial_{\xi}\pfL(\xi,s\vert x_{0}\rightarrow a)\vert_{\xi=a}} = \frac{\partial_{\xi}\pfL(\xi,s\vert a)\vert_{\xi\rightarrow a}}{\pfL(a,s\vert a)}.
\end{equation}
Finally, combining Eqs.~(\ref{ratio_pd_pf}) and (\ref{derivative-pabs-pfree}), we are led to
\begin{equation}\label{derivative-pabs-pdelta}
D\dplL = \frac{\pdL^{++}(a,s\vert x_{0})}{\pdL^{++}(a,s\vert a)}.
\end{equation}
\subsubsection*{Relation between $\prL$ and $\pfL$}
We follow Ref.~\cite{Grosche-1995} and treat reflective BC by considering potentials that are perturbed by the derivative of the $\delta$-function
\begin{equation}
W(x) = V(x) - \gamma\delta'(x-a).
\end{equation}
Analogously to the absorbing BC case, we first consider the free particle as unperturbed problem, that is $V(x)=0$, and take the limit of an infinitely repulsive perturbation $\gamma\rightarrow -\infty$. Then, one has \cite{Grosche-1995}
\begin{equation}\label{Ref-Free}
\prL\argL = \pfL\argL - \frac{\partial_{\xi}p_{\text{free}}(x,s\vert \xi)\vert_{\xi=a}\partial_{\xi}p_{\text{free}}(\xi,s\vert x_{0})\vert_{\xi=a}}{\partial_{\xi}\partial_{\phi} p_{\text{free}}(\xi,s\vert \phi)\vert_{\xi=\phi=a}}.
\end{equation}
Employing Eqs.~(\ref{derivative-chain-pfree-identities}) and (\ref{fundamental-free}), we conclude from Eq.~(\ref{Ref-Free}) that
\begin{equation}\label{Ref-Free-DFree}
\prL(a,s\vert x_{0}) = - \frac{1}{D}\frac{\pfL(a,s\vert x_{0})}{\partial_{\xi}\pfL(a,s\vert \xi)\vert_{\xi\rightarrow a}}, 
\end{equation}
and hence, using Eq.~(\ref{derivative-pabs-pfree}), we arrive at
\begin{equation}\label{pref-deriv-pabs}
\frac{\prL(a,s\vert x_{0})}{\prL(a,s\vert a)} = \frac{\pfL(a,s\vert x_{0})}{\pfL(a,s\vert a)} = D\dplL = \frac{\partial_{\xi}\pfL(\xi,s\vert x_{0})\vert_{\xi=a}}{\partial_{\xi}\pfL(\xi,s\vert x_{0}\rightarrow a)\vert_{\xi=a}}.
\end{equation}
In addition, Eq.~(\ref{Ref-Free-DFree}) tells us that
\begin{equation}\label{DRef}
\partial_{\xi}\prL(a,s\vert \xi)\vert_{\xi\rightarrow a} = - \frac{1}{D}, 
\end{equation}
Another consequence of Eq.~(\ref{Ref-Free}) is the relation
\begin{equation}\label{pref-inv-pabs}
\prL(a,s\vert a) = -\frac{1}{D^{2}}\frac{1}{\partial_{\xi}\partial_{\phi} \paL(\xi,s\vert\phi)\vert_{\xi=\phi=a}},
\end{equation} 
where we have used Eqs.~(\ref{Ref-Free-DFree}) and (\ref{second-derivative-pabs-pfree}).  
We note that Eqs.~(\ref{DRef}), (\ref{pref-inv-pabs}) and the relation
\begin{equation}
\frac{\prL(a,s\vert x_{0})}{\prL(a,s\vert a)} = D\dplL
\end{equation}
(see Eq.~(\ref{pref-deriv-pabs})) have also been obtained by other means \cite{Agmon-1991}.
\subsubsection*{Relation between $\pdL^{++}$ and $\paL$}
Now, we are sufficiently prepared to turn to the actual proof of the PDX version that relates $\paL$ with $\pdL^{++}$. To this end, we make the following construction. Starting point is the potential $W(x) = V(x) - \gamma_{\text{abs}}\delta(x-a_{\text{abs}})$, where $V(x) = -\gamma\delta(x-a)$, that is, we have a $\delta$-function potential perturbed by a second $\delta$-function potential localized at a different position. We assume that $a_{\text{abs}} < a$. Then, the propagators $\tilde{p}_{W}(x,s\vert x_{0})$ and
$\tilde{p}_{V}(x,s\vert x_{0})$ satisfy
\begin{eqnarray}\label{preAbsRad}
\tilde{p}_{W}(x, s\vert x_{0})=\tilde{p}_{V}(x, s\vert x_{0}) + \gamma_{\text{abs}}\frac{\tilde{p}_{V}(x, s\vert a_{\text{abs}})\tilde{p}_{V}(a_{\text{abs}}, s\vert x_{0})}{1-\gamma_{\text{abs}}\tilde{p}_{V}(a_{\text{abs}}, s\vert a_{\text{abs}})}.
\end{eqnarray}
Note that we have for the unperturbed GF $\tilde{p}_{V}(x, s\vert x_{0}) = \pdL^{++}\argL$, while $\tilde{p}_{W}(x, s\vert x_{0})$ is not equal to any of the GF we introduced before.  
However, as a next step, we consider the limit $a_{\text{abs}}\rightarrow a$ and $\gamma_{\text{abs}}\rightarrow-\infty$. In this limit, the potential $W(x)$ assumes the form corresponding to a purely absorbing BC, that is $W(x) = -\lim_{\gamma\rightarrow-\infty}\gamma\delta(x-a)$. Therefore, Eq.~(\ref{preAbsRad}) becomes 
\begin{equation}
\paL\argL = \pdL^{++}\argL - \frac{\pdL^{++}(x,s\vert a)\pdL^{++}(a,s\vert x_{0})}{\pdL^{++}(a, s\vert a)}.
\end{equation}
Now, the PDX formula for $\pdL^{++}$ immediately results from the just established convolution relation by virtue of Eq.~(\ref{derivative-pabs-pdelta}): 
\begin{equation}
\pdL^{++}\argL = \paL\argL + D^{2}\dprL\pdL^{++}(a,s\vert a)\dplL.
\end{equation}
\subsubsection*{Relation between $\prL$ and $\paL$}
To analyze this case we consider a potential of the form $W(x) = V(x) - \gamma_{\text{ref}}\delta^{\prime}(x-a)$, where the unperturbed potential is given by a $\delta$-function, that is, $V=-\gamma_{\text{abs}}\delta(x-a)$. We take the limit of an infinitely repulsive $\delta$- and $\delta^{\prime}$- potential. More precisely, we consider 
\begin{equation}\label{Wabsref}
\lim_{\gamma_{\text{abs}}, \gamma_{\text{ref}}\rightarrow-\infty}W(x)
\end{equation} 
and obtain
\begin{equation}
\prL\argL = \paL\argL - \frac{\partial_{\xi}\paL(x,s\vert \xi)\vert_{\xi=a}\partial_{\phi}\paL(\phi,s\vert x_{0})\vert_{\phi=a}}{\partial_{\xi}\partial_{\phi}\paL(\xi,s\vert \phi)\vert_{\xi=\phi=a}}.
\end{equation}
Then, by making use of Eq.~(\ref{pref-inv-pabs}), we obtain the PDX for $\prL\argL$
\begin{equation}\label{pref-pabs}
\prL\argL = \paL\argL + D^{2}\partial_{\xi}\paL(x,s\vert \xi)\vert_{\xi=a}\prL(a,s\vert a)\partial_{\phi}\paL(\phi,s\vert x_{0})\vert_{\phi=a}.
\end{equation}
We note that the considered potential (Eq.~(\ref{Wabsref})) also gives rise to the relation
\begin{equation}
\paL\argL = \prL\argL - \frac{\prL(x,s\vert a)\prL(a,s\vert x_{0})}{\prL(a, s\vert a)}.
\end{equation}
\subsubsection*{Relation between $\pckL$ and $\paL$}
It has been shwon that the GF $\pckL$ corresponding to a radiation BC enjoys a path integral representation, given one considers a $\delta$-function potential in the presence of a Neumann BC \cite{Clark-1980, Farhi-1990}. Put differently, this means that $\pckL$ is obtained from the potential $W(x) = V(x) + \mathcal{V}(x)$, where the unperturbed potential and the perturbation are given by
\begin{eqnarray}
V(x) &=& -\lim_{\gamma\rightarrow-\infty}\delta^{\prime}(x-a), \\
\mathcal{V}(x) &=& \kappa_{a}\delta(x-a),
\end{eqnarray}   
respectively.
Using once again the general convolution formula Eq.~(\ref{W-V-GF}), we arrive at
\begin{equation}\label{prad-pref}
\pckL\argL= \prL\argL - \kappa_{a}\frac{\prL(x,s\vert a)\prL(a,s\vert x_{0})}{1+\kappa_{a}\prL(a,s\vert a)}.
\end{equation}
Eq.~(\ref{prad-pref}) has been derived before \cite{SzaboLammWeiss:1984, Agmon-1991} in a different way.
From Eq.~(\ref{prad-pref}) we obtain
\begin{equation}\label{prad-a}
\pckL(a,s\vert x_{0}) = \frac{\prL(a,s\vert x_{0})}{1+\kappa_{a}\prL(a,s\vert a)}.
\end{equation}
Using the established identity Eq.~(\ref{prad-a}) as well as Eq.~(\ref{pref-deriv-pabs}), we obtain from Eq.~(\ref{prad-pref})
\begin{eqnarray}\label{prad-pref-2}
&&\pckL\argL= \prL\argL  \nonumber \\
&-&D^{2}\kappa_{a}\partial_{\xi}\paL(x,s\vert \xi)\vert_{\xi=a}\partial_{\phi}\paL(\phi,s\vert x_{0})\vert_{\phi=a}\pckL(a,s\vert a)\prL(a,s\vert a).
\end{eqnarray}
By virtue of Eqs.~(\ref{pref-pabs}), (\ref{prad-pref-2}) and again (\ref{prad-a}), we finally arrive at the PDX for $\pckL\argL$
\begin{equation}
\pckL\argL= \paL\argL + D^{2}\partial_{\xi}\paL(x,s\vert \xi)\vert_{\xi=a}\pckL(a,s\vert a)\partial_{\phi}\paL(\phi,s\vert x_{0})\vert_{\phi=a}.
\end{equation} 
We finally note that, due to
\begin{equation}
1-\frac{D^{2}}{\kappa_{a}}\partial_{\xi}\partial_{\phi}\paL(\xi, s\vert \phi)\vert_{\xi=\phi=a} = \frac{1}{\kappa_{a} \pckL(a,s\vert a)},
\end{equation}
see Eqs.~(\ref{pref-inv-pabs}) and (\ref{prad-a}),
the PDX for $\pckL$ gives rise to a representation of $\pckL$ in terms of $\paL$ only
\begin{equation}\label{prad-pabs}
\pckL\argL = \paL\argL + \frac{D^{2}}{\kappa_{a}}\frac{\partial_{\xi}\paL(x,s\vert \xi)\vert_{\xi=a}\partial_{\phi}\paL(\phi,s\vert x_{0})\vert_{\phi=a}}{1-\frac{D^{2}}{\kappa_{a}}\partial_{\xi}\partial_{\phi}\paL(\xi,s\vert \phi)\vert_{\xi=\phi=a}}.
\end{equation}
This shows that, dependent on which term of the potential $W(x)$ is chosen as perturbation, we obtain either Eq.~(\ref{prad-pref}) or (\ref{prad-pabs}), see also Ref.~\cite{Farhi-1990}.
\section{Appendix}
Let us define
\begin{eqnarray}
u(x,s) &\equiv& \pfL(x,s\vert x_{0}), \label{u-def} \\ 
v(x,s) &\equiv& \pfL(x,s\vert a). \label{v-def}
\end{eqnarray}

Then, $u(x,s)$and $v(x,s)$ satisfy the free-space diffusion equation, 
\begin{eqnarray}
su(x,s) - \delta(x-x_{0}) &=& D\frac{\partial^{2}u(x,s)}{\partial x^{2}}, \label{u-DE}  \\ 
sv(x,s) - \delta(x-a) &=& D\frac{\partial^{2}v(x,s)}{\partial x^{2}}.  \label{v-DE} 
\end{eqnarray}
Next, assuming $x_{0} > a$, multiplying Eq.~(\ref{u-DE}) with $v(x,s)$ and Eq.~(\ref{v-DE}) with $u(x,s)$, as well as substracting the resulting equations and integrating the outcome from $a$ to $\infty$ leads to
\begin{equation}\label{fundamental-free}
\pfL(a,s\vert x_{0}) = D\bigg[\pfL(a,s\vert a) \dpflL -\, \pfL(a,s\vert x_{0})\frac{\partial\pfL(\xi,s\vert a)}{\partial\xi}\bigg\vert_{\xi\rightarrow a}\bigg],
\end{equation}
where we have used the definitions Eqs.~(\ref{u-def}) and ~(\ref{v-def}), the BC at infinity, $\lim_{x\rightarrow\infty}\pfL\argL\rightarrow 0$, and $\pfL\argL = \pfL(x_{0},s\vert x)$.
Taking the limit $x_{0}\rightarrow a$, it follows that
\begin{equation}\label{commutator-free}
\partial_{\xi}\pfL(\xi,s\vert x_{0}\rightarrow a)\vert_{\xi=a} - \partial_{\xi}\pfL(\xi,s\vert a)\vert_{\xi\rightarrow a} = D^{-1}.
\end{equation}
Furthermore, dividing Eq.~(\ref{fundamental-free}) by $\pfL(a,s\vert a)$ and using Eq.~(\ref{commutator-free}), we obtain the chain of identities
\begin{equation}\label{chain-pfree-identities}
\frac{\pfL(a,s\vert x_{0})}{\pfL(a,s\vert a)} = \frac{D\partial_{\xi}\pfL(\xi,s\vert x_{0})\vert_{\xi=a}}{1+D\partial_{\xi}\pfL(\xi,s\vert a)\vert_{\xi\rightarrow a}}  = \frac{\partial_{\xi}\pfL(\xi,s\vert x_{0})\vert_{\xi=a}}{\partial_{\xi}\pfL(\xi,s\vert x_{0}\rightarrow a)\vert_{\xi=a}}. 
\end{equation}
Finally, as a consequence of Eq.~(\ref{chain-pfree-identities}), we arrive at
\begin{equation}\label{derivative-chain-pfree-identities}
\frac{\partial_{\xi}\partial_{\phi}\pfL(\xi,s\vert \phi)\vert_{\xi=\phi=a}}{\partial_{\xi}\pfL(\xi,s\vert x_{0}\rightarrow a)\vert_{\xi=a}} = \frac{\partial_{\xi}\pfL(\xi,s\vert a )\vert_{\xi\rightarrow a}}{\pfL(a,s\vert a)}.
\end{equation}
\section*{Acknowledgments}
This research was supported by the Intramural Research Program of the NIH, National Institute of Allergy and Infectious Diseases. 

\end{document}